\documentclass[aps,pre,twocolumn,showpacs,floatfix]{revtex4}
\usepackage{epsfig}

\newcommand{\Eqn}[1]{Eq.~(\ref{#1})}     
     
\newcommand{\Sec}[1]{Section~\ref{#1}}     
     
\newcommand{\Fig}[1]{Fig.~\ref{#1}}

\begin{document}
\title{Phase Transition in Liquid Drop Fragmentation} \author{Cristian
  F.~Moukarzel~\footnote{Corresponding author: cristian@mda.cinvestav.mx},
  Silvia F.~Fern\'andez-Sabido~\footnote{Present address: Laboratoire
    Informatique d'Avignon, France.} and J.~C.~Ruiz-Su\'arez~\footnote{Present
    address: Cinvestav-Monterrey, Av.~Cerro de las Mitras 2565, Monterrey NL,
    64060, M\'exico}}
\affiliation{CINVESTAV del IPN Unidad M\'erida, Depto.\ de F\'\i sica Aplicada, \\
  97310 M\'erida, Yucat\'an, M\'exico } \date{\today}
\begin{abstract}
  A liquid droplet is fragmented by a sudden pressurized-gas blow, and the
  resulting droplets, adhered to the window of a flatbed scanner, are counted
  and sized by computerized means. The use of a scanner plus image recognition
  software enables us to automatically count and size up to tens of thousands
  of tiny droplets with a smallest detectable volume of approximately 0.02 nl.
  Upon varying the gas pressure, a critical value is found where the
  size-distribution becomes a pure power-law, a fact that is indicative of a
  phase transition. Away from this transition, the resulting size
  distributions are well described by Fisher's model at coexistence. It is
  found that the sign of the surface correction term changes sign, and the
  apparent power-law exponent $\tau$ has a steep minimum, at criticality, as
  previously reported in Nuclear Multifragmentation
  studies~\cite{PCTEFA84,MBCEOP88}.  We argue that the observed transition is
  not percolative, and introduce the concept of dominance in order to
  characterize it.  The dominance probability is found to go to zero sharply
  at the transition. Simple arguments suggest that the correlation length
  exponent is $\nu=1/2$. The sizes of the largest and average fragments, on
  the other hand, do not go to zero but behave in a way that appears to be
  consistent with recent predictions of Ashurst and
  Holian~\cite{AHDFB99,AHDSD99}.
\end{abstract}
\maketitle
\section{Introduction}
\label{sec:introduction}
Fragmentation processes are important in several different fields, including
powder technology~\cite{PFROA05,DTGPS04}, soil
physics~\cite{RSFFS91,PFMF97,ESDA03,MGAOTF03} nuclear
physics~\cite{RWMMA01,DD-GLTTM05}, astrophysics~\cite{BKGFOT83,OMBSAO98},
genetics~\cite{MLMOF91,MCDI91}, sprays~\cite{VMDLSF04,MVOSF04,SRWCAO99}, fuel
combustion~\cite{SLBSID94,SGTTTS98}, brittle
fracture~\cite{WKHBOS05,WKHFOS04,ASMO06} and geology~\cite{XLWPDO02,JRFDA03},
just to name a few.
\\
Because of the nature of the fragmentation process, only the final
distribution of fragment sizes is usually observable, thus becoming the main
quantity of interest.  Size-distributions have been characterized by different
empirical and theoretical statistical laws in the past.  The lognormal
distribution can be expected whenever a breakdown process can be described as
a multiplicative cascade, and has been applied in many fragmentation
examples~\cite{ASMO06,SGTTTS98,DLBASS96,BSMSAF96,IMFOL92}.  Simmons'
law~\cite{CHFTPO97,HFDPA93,HFNDD92} appears to hold universally for sprays,
and states that the mass distribution is Gaussian when written in terms of the
square root of drop diameters.  Villermaux~\cite{VMDLSF04,VUIO04,MVOSF04}
discusses the breakup process of ligaments in liquid jets, and concludes that
each ligament gives rise to a distribution of droplet sizes that is
essentially a Gamma distribution.  The Weibull distribution can be derived
from first-principles for the fragmentation of solids by
cracking~\cite{BWDOT95,ZGPATW99}, and reads \hbox{$ \%N(>m) = e^{-(m/\hat
    m)^{q}}$,} where $\%N(>m)$ is the fraction of droplets with mass larger
than $m$.  Brown and Wohletz~\cite{BWDOT95} remark that the Weibull
distribution may look similar to the lognormal, and suggest that the apparent
goodness of fits using lognormal could be fortuitous in some cases.  Other
authors have used the (empirical) {\it Rosin-Rammler}
form~\cite{BWDOT95,DTGPS04,ALBCOF97}, which applies to the accumulated mass
density in clusters of diameter larger than $D$, and states that
\hbox{$\%M(>D) = e^{-(D/D_m)^{q}}$}.
\\
In contrast with these approaches, which postulate exponential, or else
rapidly decaying size distributions, Oddershede et al~\cite{ODBSCI93} have
claimed that the size-distribution of fractured objects is an inverse
power-law, and furthermore suggested that a form of self-organized criticality
(SOC)~\cite{BTWSC88} might be responsible for this.  Power-law distributions
are often found right at second order phase transitions, in which case one or
more parameters must take precisely defined values.  However there are systems
whose dynamics converges to a stationary state that is critical, a phenomenon
termed SOC~\cite{BTWSC88} and for which no parameter tuning is needed.  Later
investigations~\cite{MBCPL96,TRMOF99,SRRDCI00,TSMOF00,OPLDDM00,OLCFEA07}
report ``composite'' power-laws, or transitions between lognormal and
power-laws as the fragmentation energy is
increased~\cite{IMFOL92,SMLCID96,SGTTTS98}.
\\
Recent works~\cite{KHTFD99,AHTUIF00,KSHSOI03,WKHBOS05,HWKF06} have reported
that fragmentation processes are only critical (and the ensuing
size-distribution a power-law) at a precisely defined value of the impact
energy. Within this picture, the fragmentation phenomenon goes through a
nonequilibrium phase transition at some well-defined value of the impact
energy. Evidences of criticality at a well defined energy where also found in
nuclear multifragmentation (NMF)~\cite{RWMMA01,DD-GLTTM05} experiments, and
have been interpreted in the context of a percolative
transition~\cite{CMNB86}, or else in relation with liquid-gas
coexistence~\cite{FCRFYA85,MBCEOP88,PMRPTN95,BLBCEO95,MPGPCI96,PD-GGFOP98,GFFPTL01,EMPLTV02,DBGESO04}
in nuclear matter.
\\
In this work, water-glycerine drops are broken by a controlled-pressure gas
blow, and the resulting droplets are counted and sized by computerized methods
that involve scanning and image-processing. The only control parameter, the
gas pressure, varies between 11 and 100 psi, and this in turn determines the
amount of energy available for the breakdown event.  Various statistical
measures characterizing the resulting size-distributions are obtained and
analyzed, for a minimum of ten and a maximum of fifty breakdown events at each
pressure. Our main finding is the existence of a phase transition around
\hbox{$P_c \approx 17 \hbox{psi}$}, where the size-distribution becomes a pure
power-law with an exponent $\tau_{min} \approx 1$ within a limited size-range.
Away from $P_c$, the size distribution is well described by the product of a
power-law and an exponentially decaying term (\Eqn{eq:14}), as given by
Fisher's model at coexistence~\cite{FTOC67}.
\\
The rest of this article is organized as follows. In
\Sec{sec:size-distributions} a brief discussion is done of some theoretical
approaches which have been used to characterize size-distributions in the
past.  \Sec{sec:exper-setup-data} describes our experimental apparatus, as
well as the data processing techniques we use to obtain statistics on size
distributions. Our results are presented in \Sec{sec:results}. In
\Sec{sec:size-distribution} the resulting size distributions are characterized
by means of a simplified Fisher expression (\Eqn{eq:14}), essentially a power
law with an exponential cutoff.  The total surface is discussed in
\Sec{sec:excess-surface}. \Sec{sec:observ-trans-perc} discusses whether the
observed transition can be characterized as percolative. The mass of the
largest fragment (\Sec{sec:largest-fragment} ) and moment correlations
(\Sec{sec:moment-correlations}) do not support such interpretation.
\Sec{sec:ashurst-holi-theory} discusses the behavior of the largest and
average mass fragment in the context of recent theoretical
elaborations~\cite{AHDFB99,AHDSD99}. In \Sec{sec:domin-prob} the concept of
dominance is introduced and applied to our data.  Finally, our results are
discussed in \Sec{sec:discussion}.
\section{Size distributions}
\label{sec:size-distributions}
In recent years, several brittle fragmentation~\cite{WKHFOS04,WKHBOS05,HWKF06}
and NMF~\cite{CMNB86,RWMMA01} experiments have been interpreted in the context
of Percolation Theory~\cite{SAITP94}. A percolation transition is claimed to
exist between a ``connected'' phase and ``disconnected'' phase, as the impact
energy is increased. In the connected phase, a large cluster (drop, fragment)
exists after the fragmentation event, that contains a significant fraction of
the total mass.  In the disconnected phase no such large cluster exists. The
natural order parameter for a percolative transition of this type is the
average size of the largest fragment. Measurements of this order
parameter~\cite{CMNB86,WKHFOS04} have shown that it goes to nearly zero at and
above the transition, in a way that is consistent with the proposed
interpretation.
\\
At the percolation critical point, the number distribution $N(x)$ of clusters
of mass $x$ becomes a power-law $N(x) \sim x^{-\tau}$, where the exponent
$\tau$ takes values close to $2.3$ in three dimensions.  NMF experiments have
been reported to give results consistent with $\tau \approx
2.3$~\cite{CMNB86,RWMMA01} as well, while brittle fragmentation experiments
seem to produce different values of $\tau$.
\\
NMF experiments are sometimes more generally
discussed~\cite{MNWCBI05,DBGCBI03,RBCEO01,EMPNMP00,SRLNMP97,FBKSOP96,BLBCEO95}
in the context of Fisher's Droplet Model~\cite{FTOC67}, which states that the
number of clusters (droplets, nuclei) of mass $x$ is given by
\begin{equation}
  N(x) = N_0 x^{-\tau} e^{B(p,T)x - C(p,T) x^{\sigma}},
  \label{eq:11}
\end{equation}
where $\tau$ and $\sigma$ are independent critical exponents.  Coefficients
$B(p,T)$ and $C(p,T)$ depend on surface tension, temperature, and chemical
potentials in the ``liquid'' and ``gas'' phases:
\begin{equation}
  B(p,T) = \frac{\Delta \mu}{T} = \frac{\mu_{gas} - \mu_{liq}}{T},
  \label{eq:12}
\end{equation}
and 
\begin{equation}
  C(p,T) = \frac{c_0 \epsilon}{T}.
  \label{eq:13}
\end{equation}
Here $c_0$ is a constant that depends on surface tension, and
$\epsilon=(T_c-T)/T_c$ measures the departure from the critical temperature.
\\
The term $B(p,T) x$ measures the bulk contribution to the free energy of a
droplet of size $x$.  In a $(p,T)$ phase diagram, $B<0$ if the ``Gas'' phase
is thermodynamically stable, i.e. its chemical potential is lower than that of
the ``cluster'', or ``liquid'' phase.
\\
The term $C x^{\sigma}$ measures the surface energy of a droplet of size $x$,
so that $\sigma$ is the critical exponent that relates the dimension of the
surface to that of the volume of a droplet. For compact droplets in 3d one has
$\sigma=2/3$. For the liquid-gas transition $\sigma$ takes values close to
$2/3$.
\\
On the coexistence line, $\mu_l = \mu_g$ and thus $B=0$, so
\begin{equation}
  N(x) = N_0 x^{-\tau} e^{- C x^{\sigma}}.
  \label{eq:14}
\end{equation}
Furthermore, at the critical point one has $C=0$, and the size-distribution
becomes a pure power-law with exponent $\tau$. 
\section{Experimental Setup and Data Acquisition}
\label{sec:exper-setup-data}
\subsection{Setup}
\label{sec:setup}
\begin{figure}[h!]
  \centering
  \epsfig{figure=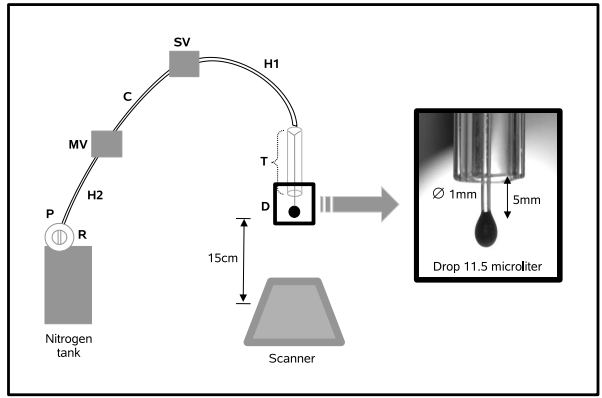,width=8cm,angle=0}
  \caption{Experimental Setup }
  \label{fig:setup}
\end{figure}

Our experimental setup is schematically shown in \Fig{fig:setup}.  An open,
straight glass tube (T) of 6 cm length and 7 mm inner diameter, has a solid
coaxial glass rod (1mm diameter) attached to its inner surface, with the rod's
tip protruding 5 mm downwards from the lower end of the tube.  From this tip
hangs a small (11.5 $\mu l$~\footnote{This is about the maximum volume that
  can be held by the liquid's surface tension.}) liquid drop (D) of an opaque
solution obtained by mixing equal quantities of glycerol and water, plus a
small amount of water-soluble black dye. The upper end of this glass tube is
connected to a pressurized chamber (C) via a 40cm length plastic (H1) hose,
and isolated from it by a solenoid valve (SV).  The tube is fixed so that the
droplet hangs 15 cm above the glass window of a flatbed scanner.
\\
A nitrogen cylinder is connected to the chamber through a plastic hose (H2),
and a pressure regulator (R), which is provided with a manometer (P). The
chamber is formed by a 10 cm length, 3/8 inch inner diameter plastic hose and
is limited by two bypass valves, one manual (MV), the other electromagnetic
(SV).
\\
For each droplet breakdown experiment, the nitrogen pressure was selected with
regulator R, and the manual valve was opened until the pressure in chamber C
equilibrated, after which it was closed again.  Then, the electromagnetic
valve was opened by means of an electric switch, releasing the gas contained
in chamber C, which then flowed through tube T, fragmenting the droplet and
projecting its fragments downwards.  The resulting droplets, adhered to the
scanner's glass window, were scanned and subsequently counted and sized with
the help of specialized computer software.
\\
Before each breakdown experiment, the scanner glass was treated with a
hydrophobic solution containing Teflon.  Because of this treatment of the
glass surface, the contact angle of the liquid onto the glass was close to 90
degrees, so that the droplets turned out to be very similar in shape to
semi-spheres. After the fragmentation, a thin spacing was inserted between the
scanner lid and its glass window, and a high-resolution (600 dpi) scan was
performed.  Several (from 10 to 50) fragmentation experiments were carried out
varying the jet pressure P, from 11 psi to 100 psi.  The number of droplets
obtained varied, from a few ones at 11 psi to tens of thousands at 100 psi.
\subsection{Data Processing}
\label{sec:data-processing}
After each fragmentation, a scanning operation was performed.  All scanned
images were obtained at 600 dpi-grayscale, and stored using a lossy compressed
graphic format, for later processing.  Because of the non-negligible height of
the droplets, most images contained faint shadows, which were removed by image
thresholding~\footnote{Using Adobe Photoshop}.  Subsequent image processing
and droplet identification, counting, and sizing, was performed with the help
of ImageJ~\footnote{http://rsb.info.nih.gov/ij/}
\\
\begin{figure}[h!]
  \centering
  \psfig{figure=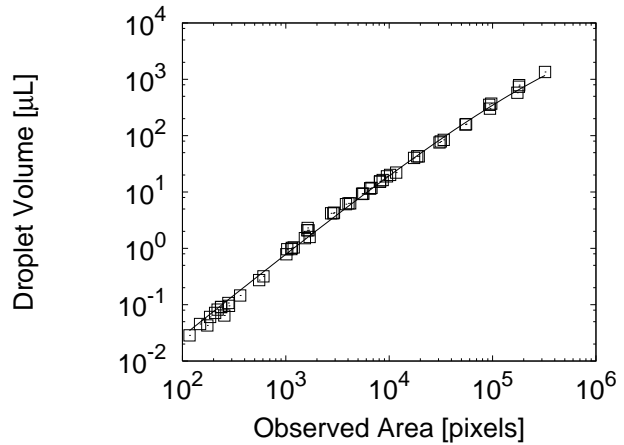,width=6.0cm,angle=270}
  \caption{Data used for fitting \Eqn{eq:avolrelation} (squares) and the
  resulting fit (solid line). }
  \label{fig:avolfit}
\end{figure}
The output from ImageJ consisted essentially of a set of droplet areas, in
pixels.  In order to estimate droplet volumes $V_i$ from the observed areas
$\Omega_i$, we did the following. If gravity effects were negligible then all
droplets would be plane sections of a sphere and their volumes $V_i$ would be
given by
\begin{equation}
  V_i = \lambda_1 \Omega_i^{3/2},
\label{eq:vfromarea0}
\end{equation}
with $\lambda_1$ some constant. However (\ref{eq:vfromarea0}) is a good
approximation only for the tiniest droplets, whose shapes are less affected by
gravity. Larger droplets suffer a significant flattening and therefore
\Eqn{eq:vfromarea0} must be corrected.
\\
Under the effect of gravity, the height $h(\Omega)$ of a liquid droplet on a
flat horizontal surface is proportional to $\sqrt{\Omega}$ when $\Omega$ is
small, but saturates to a constant value $h_\infty$ for large values of
$\Omega$, in which case the fluid forms a flat ``lake'' of approximately
constant height. Thus $V$ satisfies (\ref{eq:vfromarea0}) for small $\Omega$,
but is proportional to $\Omega$ for large $\Omega$. Assuming, for simplicity,
a stretched exponential crossover, we write
\begin{equation}
  V(\Omega) = 
    \lambda_1 \Omega^{3/2} e^{-(\Omega/\Omega_0)^\beta} +
    \lambda_2  \Omega \left [ 1-e^{-(\Omega/\Omega_0)^\beta} \right ],
\label{eq:avolrelation}
\end{equation}
where $\lambda_1, \lambda_2, \Omega_0$, and $\beta$ are to be obtained from
fits of calibration data (See \Fig{fig:avolfit}). The resulting values where 
$\lambda_1=3.15 \times 10^{-5}$, $\lambda_2=8.3\times 10^{-7}$ $\Omega_0=10^5$
pixels, and $\beta = 0.366$.  In order to obtain the data required for the
fit, the projected areas were measured for several droplets having known
volumes.  Their volumes were obtained using calibrated pipettes (for the
larger ones), and glass capillaries (for the smaller ones).  The data used for
the fit, together with the resulting fit, are displayed in \Fig{fig:avolfit}.
Our final expression for the volume/area relationship extrapolates to \hbox{$V
  \sim 2 \times 10^{-5} \mu$l} when \hbox{$\Omega =$ 1 pixel}, implying that
the smallest detectable volume is \hbox{$V_{min}=$0.02 nl}.
\begin{figure}[h!]
  \centering
  \psfig{figure=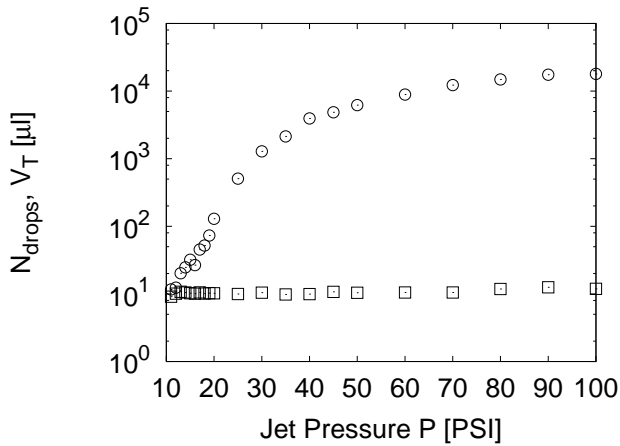,width=6.0cm,angle=270}
  \caption{Average total observed volume $V_T$ (squares) and
    average number of droplets $N_{drops}$ (circles) as a function of jet
    pressure $P$.  The initial volume of the droplet is 11.5 $\mu$l.}
  \label{fig:avgVN}
\end{figure}
\Eqn{eq:avolrelation}, together with the parameter values obtained from the
fit, were embedded into a C program, which read droplet areas (as calculated
by ImageJ), and used this expression to calculate volumes from them,
additionally implementing the statistical analysis of size distributions to be
described in \Sec{sec:results}.
\section{Results}
\label{sec:results}
\begin{figure}[h!]
  \centering
  \psfig{figure=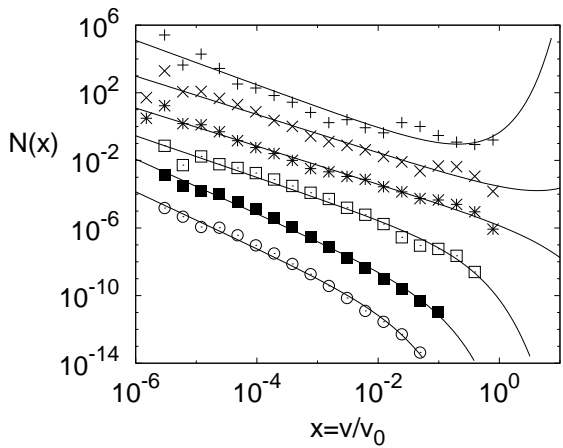,width=6cm,angle=270}    
  \caption{$N(x)$, the number of  drops of relative mass $x=V/V_0$, for (from
  top to bottom)
  $P=$11, 15, 17, 25, 60, and 100 PSI. Solid
  lines are fits of \Eqn{eq:14} with $\sigma=2/3$. Plots have been displaced
  downwards for clarity. } 
  \label{fig:Ndens.in.x}
\end{figure}
For each breakdown event, a set of droplet volumes $V_i$, $i=1,2,\ldots,n$ was
obtained as described in \Sec{sec:data-processing}, and the moments $M_k =
\sum_i x_i^k$ were calculated in terms of \hbox{$x_i=V_i/\sum_{j=1}^n V_j$},
the droplet volume fractions. The average number of droplets, as well as the
average total volume $V_T =\sum_i V_i$, are displayed in \Fig{fig:avgVN} as a
function of jet pressure P. The fact that $V_T$ remains almost constant and
equal to the initial volume confirms that most of the mass resulting from each
fragmentation gets caught by the scanner window, i.e. that there are no
significant mass losses, even at the largest pressures considered.
\subsection{Size distributions}
\label{sec:size-distribution}
\begin{figure}[h!]
  \centering
  \psfig{figure=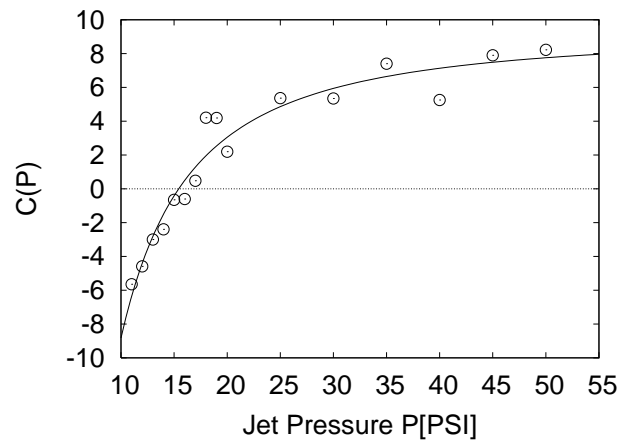,width=6.0cm,angle=270}        
  \caption{Amplitude $C(P)$ (circles) of surface correction term in \Eqn{eq:14}, for
    $\sigma=2/3$. The solid line is an empirical fit (see text).}
  \label{fig:coeff}
\end{figure}

\begin{figure}[h!]
  \centering
  \psfig{figure=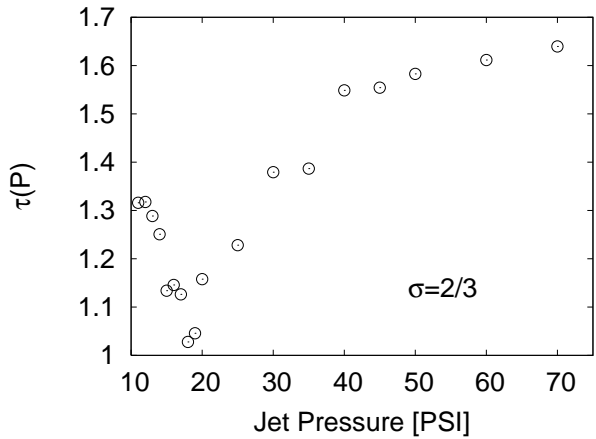,width=6.0cm,angle=270}
  \caption{Fits of Fisher model (\Eqn{eq:14}) to the droplet density $N(x)$ (\Fig{fig:Ndens.in.x}),
    produce an apparent exponent $\tau$ (circles) showing a deep minimum
    \hbox{$\tau_{min} \approx 1$} around 18 PSI.}
  \label{fig:effexp}
\end{figure}

\Fig{fig:Ndens.in.x} shows log-log plots of the normalized fraction of
droplets $N(x)$ with volume fraction $x$, for several values of the breakup
pressure $P$. Continuous lines in \Fig{fig:Ndens.in.x} are the result of
fitting Fisher's Model in the form indicated by \Eqn{eq:14}, which is
appropriate for the coexistence region.  We fixed $\sigma=2/3$ and adjusted
$N_0(P)$ (not shown), $C(P)$ (\Fig{fig:coeff}) and $\tau(P)$ (the
\emph{apparent} exponent, see \Fig{fig:effexp}) independently for each
pressure $P$. We checked that the qualitative features of these fits did not
change much for $\sigma$ ranging from $1/3$ to $2$.
\\
\\
The ``surface'' correction term $C(P)$ (\Fig{fig:coeff}) changes sign around
$P_0=17$ PSI, implying that the size distribution is similar to a pure
power-law there, and giving the first hint for the existence of a phase
transition. The resulting curvature change that is apparent in
\Fig{fig:Ndens.in.x} has been reported and discussed in NMF experiments as
well~\cite{GFFPTL01,BLBCEO95}. The solid line in \Fig{fig:coeff} is a fit of
the form
\begin{equation}
  C(P) = a \left [ \left ( \frac{P}{P_0} \right )^{\mu} -1 \right ]
\end{equation}
and results in $P_0=15.5$ and $\mu=-1.5$. This suggests that 
\begin{equation}
C \sim (P-P_0)  
\label{eq:Clin}
\end{equation}
near the crossing point $P_0$.
\\
The exponential term in (\ref{eq:14}) can be written, as is usually done in
the scaling description of critical phenomena, as
\begin{equation}
e^{-C x^\sigma} = F_{\pm}( (x/x_0)),
\end{equation}
where the subscript $\pm$ means $P>P_0$ or $P<P_0$ respectively. The
``cutoff'' $x_0$ satisfies
\begin{equation}
x_0= |C|^{-1/\sigma}.
\label{eq:x0}
\end{equation}
For compact droplets we can furthermore write
\begin{equation}
x_0 = \xi^3 
\label{eq:xi}
\end{equation}
in three dimensions, where $\xi$ is a correlation length (the linear size of
the typical droplet). Considering (\ref{eq:Clin}), (\ref{eq:x0}), and
(\ref{eq:xi}) one finds that, near $P_0$,
\begin{equation}
  \xi \sim |P-P_0|^{-\nu}
\end{equation}
with $\nu =1/2$, i.e. a classical value for a thermal critical exponent. Of
course, this specific result ($\nu=1/2$) depends on the validity of
(\ref{eq:Clin}), which in view of the quality of our data cannot be taken for
granted.
\\
The apparent exponent $\tau(P)$ (\Fig{fig:effexp}) goes through a steep
minimum $\tau_{min} \approx 1$ around 18 PSI. A similar minimum in $\tau$ has
been taken as an indicator for the location of a phase transition in NMF
experiments~\cite{PCTEFA84,MBCEOP88} in the past.  In fact, Richert and
Wagner~\cite{RWMMA01} \emph{define} the critical temperature $T_c$ as the one
that minimizes $\tau(T)$. We may thus consider this steep minimum as a further
indication of a phase transition in the breakdown process, occurring somewhere
in the range 16-18 PSI for our specific experimental setup.
\subsection{Excess surface}
\label{sec:excess-surface}
\begin{figure}[h!]
  \centering
  \psfig{figure=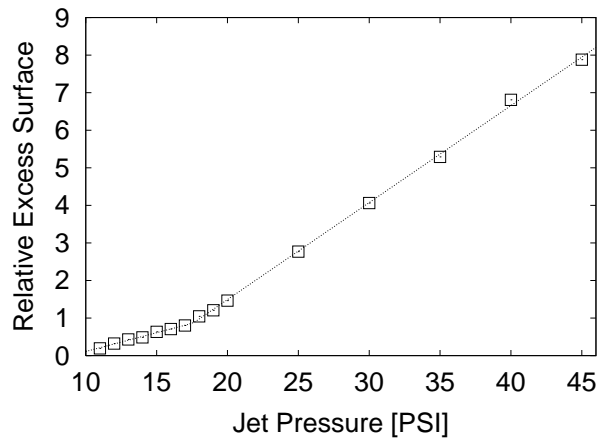,width=6.0cm,angle=270}
  \caption{Surface increase versus Pressure  has a discontinuous derivative
    around 17.6 PSI. The dotted line is a linear fit, with the location of the
    kink as a fitting parameter.}
  \label{fig:surfincrease}
\end{figure}

The total normalized surface $\Omega_F=\sum_i x_i^{2/3}$ of the final set of
drops in each breakdown process gives an estimation of the total amount of
energy transferred to the system of drops (i.e. neglecting kinetic energy) and
is thus an interesting physical quantity in the breakdown experiment. We
define the relative excess surface as
\begin{equation}
  E(P) = \frac{\Omega_F - \Omega_0}{\Omega_0}
\end{equation}
where $\Omega_0=1$ is the initial surface. By taking averages over breakdown
events at pressure $P$, the results displayed in \Fig{fig:surfincrease} are
obtained. It can be seen in this figure that the pressure-dependence is
approximately linear everywhere, but there is a sharp discontinuity in its
slope, again happening around 17 PSI. Fitting two straight lines intersecting
at $P_0$ to the data shown in \Fig{fig:surfincrease} (dotted line) one obtains
the estimate $P_0=17.6$\ PSI.
\\
If the excess surface is interpreted as an ``energy'' and the jet pressure as
a measure of the effective ``temperature'', the discontinuous slope in
\Fig{fig:surfincrease} is analogous to a discontinuity in the specific heat.
\\
\subsection{Is the observed transition percolative?}
\label{sec:observ-trans-perc}
\subsubsection{Largest fragment}
\label{sec:largest-fragment}
Recently there have been suggestions that brittle
fragmentation~\cite{WKHFOS04,WKHBOS05,HWKF06} and NMF~\cite{CMNB86,RWMMA01}
experiments can be rationalized in the context of a percolative~\cite{SAITP94}
phase transition. The validity of such interpretations rests, to a great
extent, upon the identification of a relevant order parameter which has the
properties of the infinite-cluster density in percolation. Within this
picture, the connected or percolating phase must be characterized by the
existence of macroscopic fragments, while no such fragments must exist in the
disconnected phase (at large breakdown energies). The average size of the
largest fragment, which goes to zero \emph{abruptly} at the critical energy,
plays the role of an order parameter in brittle
fragmentation~\cite{WKHFOS04,WKHBOS05,HWKF06}, so it is natural to first
consider the largest droplet size as a tentative order parameter also in the
case of our experiments.
\\
\begin{figure}[h!]
  \centering \psfig{figure=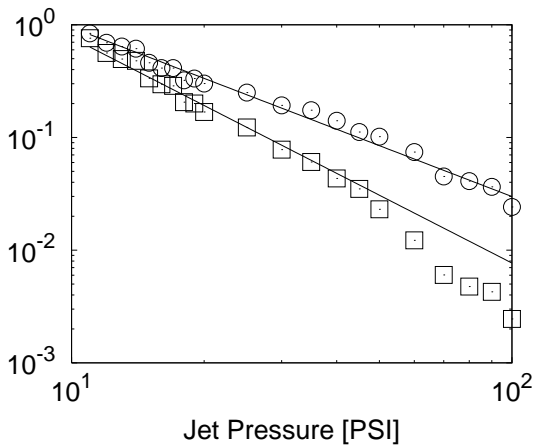,width=6.0cm,angle=270}
  \caption{ Maximum droplet size $Z_{max}$ (circles) and average droplet size
    $Z_{avg}=M_2$ (squares), as a function of jet pressure $P$. Straight lines
    are of the form $c_1 P^{-2}$ and $c_2 P^{-3/2}$ respectively.  Assuming
    that the system's \emph{expansion speed} $\eta$ is proportional to the jet
    pressure $P$, the observed dependencies $Z_{max} \sim P^{-3/2}$, and
    $Z_{avg} =M_2 \sim P^{-2}$ are consistent with recent predictions of a
    simple energy-balance theory of Ashurst and
    Holian~\cite{AHTUIF00,AHDFB99,AHDSD99}.}
  \label{fig:holianpred}
\end{figure}
The average mass $Z_{max}$ of the largest fragment is shown in
\Fig{fig:holianpred} from our droplet fragmentation experiments.  It turns out
that $Z_{max}(P)$ has a smooth pressure dependence of the type $Z_{max}(P)
\sim P^{-3/2}$, without any traces of a singular behavior. Notice that, if
there where a percolative transition, the mass of the largest droplet (the
infinite cluster) should go to zero at the critical point. Therefore we
conclude that the behavior of $Z_{max}$ does not support the existence of a
percolative transition in this system.
\subsubsection{Moment correlations}
\label{sec:moment-correlations}
In his discussion of NMF experiments, Campi~\cite{CMNB86} notices that
correlations among some moments of the size distribution are similar to those
found in Percolation. In particular, a plot is presented of $Z_{max}$ versus
$S_2=M_2/M_1$ (moments are calculated excluding the contribution of the
largest cluster) and it is argued that the rounded ``nose'' of events with
large values of $S_2$ corresponds to the percolative ``critical region''.
\\
\begin{figure}[h!]
  \centering
  \psfig{figure=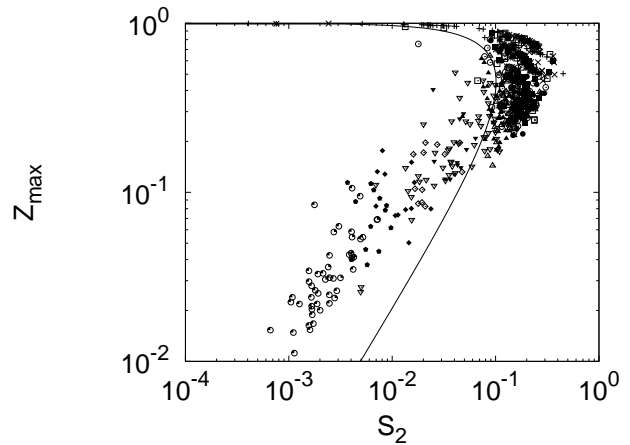,width=6.0cm,angle=270}
  \caption{Inclusive (all pressures) scatter plot of the mass $Z_{max}$ of the
    largest piece versus $S_2=M_2/M_1$, where the moments have been calculated
    without the contribution of the largest piece. Points near the ``nose'' of
    events with large values of $S_2$ correspond to pressures in the range
    16-18 PSI. Notice, however, that the existence of this feature is by
    itself not indicative of critical behavior, since it is there even for the
    trival partition described in the text (solid line).}
  \label{fig:ZmaxvsS2}
\end{figure}
\Fig{fig:ZmaxvsS2} shows a scatter plot of $Z_{max}$, the largest cluster's
mass, versus $S_2$, for each fragmentation event in our experiments. This plot
is similar in many aspects to the one considered by Campi~\cite{CMNB86}, in
particular with respect to the existence of a ``nose'' of events with large
values of $S_2$. It can be easily seen, however, that this feature by itself
is not indicative of critical behavior. In order to see this, consider the
breakup of a droplet of unitary volume into an infinite succession of droplets
$i=0,1,2,\ldots$ of volume $v_i=(1-\lambda) \lambda^i$, for $0<\lambda<1$.  As
$\lambda$ goes from almost zero to almost one, the resulting size distribution
changes from being dominated by essentially one large drop, to being composed
of a ``mist'' of tiny droplets. Clearly there is no phase transition in this
system.
\\
For this simple partition, the largest mass is $Z=v_0=(1-\lambda)$.
Furthermore, it is straightforward to show that $S_2=Z(1-Z)/(2+Z)$. A plot of
$S_2(Z_{max})$ for this simple partition is displayed in \Fig{fig:ZmaxvsS2}
(thick line). It can be seen that its overall features are the same as
reported by Campi~\cite{CMNB86}, i.e. $S_2$ goes through a maximum and then
decreases, as $Z_{max}$ decreases ($\lambda$ increases). Thus we conclude that
the rounded ``nose'' in these plots is not by itself indicative of a phase
transition, and cannot be taken as evidence of percolative behavior.
\subsection{Ashurst and Holian theory}
\label{sec:ashurst-holi-theory}
Recently Ashurst and Holian (AH)~\cite{AHDFB99,AHDSD99} (see also \"Astrom,
Holian and Timonen \cite{AHTUIF00}), considering a $d$-dimensional fluid
system that expands uniformly at rate $\eta$, predicted that the
\emph{maximum} droplet size $Z_{max}$ should behave as $\eta^{-d/2}$ while the
\emph{average} droplet size $Z_{avg}$ should decrease as $\eta^{-2d/3}$ with
increasing expansion rate~\footnote{The average considered here to is not the
  \emph{number average}
  $1/N_{drops}$ but the \emph{mass average} $\sum_i x_i^2/\sum_i x_i = M_2$.}.\\
Since in our experiments the expansion rate $\eta$ is not a control parameter,
the applicability of AH results is not obvious.  However we find (See
\Fig{fig:holianpred}) that $Z_{max} \sim P^{-1.5}$ and $Z_{avg} \sim P^{-2}$,
which is consistent with AH predictions in 3 dimensions, if the (unknown)
expansion rate $\eta$ is \emph{assumed} to be proportional to the jet pressure
$P$.
\subsection{Dominance Probability}
\label{sec:domin-prob}
\begin{figure}[h!]
  \centering
  \psfig{figure=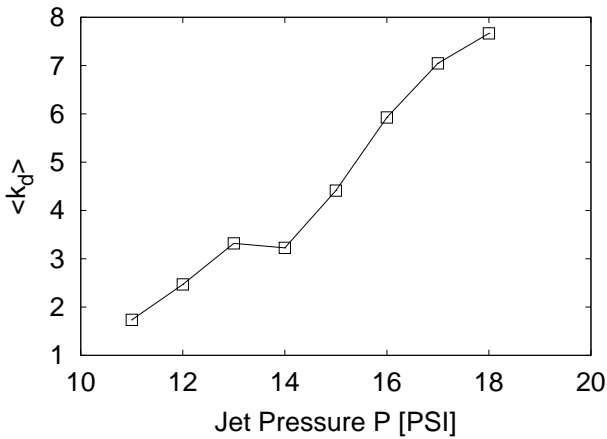,width=6.0cm,angle=270}
  \caption{Average number of dominant drops, versus pressure. For pressures
    larger than 18PSI, no dominance was ever detected.}
  \label{fig:avgk}
\end{figure}

We now set to try and identify the nature of the phase transition at $17$ PSI,
by defining a useful order parameter that we call \emph{dominance
  probability}.  Although, as seen already, the largest fragment size is a
smooth function of $P$, we have noticed that, at low pressures, most of the
mass is concentrated in a few large fragments, while at larger pressures the
mass is evenly distributed among a large number of small droplets.  More
precisely, at very low pressures the mass distribution appears to be composed
of one large fragment plus many tiny droplets. Upon increasing the jet
pressure $P$, often two large droplets appear, next events with three large
fragments become prevalent, and so on. At large enough pressures, however, the
original droplet mass is more evenly distributed among a large number of
droplets, without a clear size distinction between large and small ones.  We
then seek to define an order parameter that quantifies the property that the
mass distribution be ``dominated'' by a few large fragments or not.
\\
For an arbitrarily defined ``dominance factor'' $\gamma$ taking values
slightly smaller than 1 (one can take e.g.~$\gamma=0.9$ to exemplify the
ideas), we say that there is \emph{dominance} at level 1 if the largest
fragment's mass is larger than $\gamma$ times the total mass.  If this
condition is not satisfied, the largest fragment is removed and we check
whether the second largest fragment's mass is larger than $\gamma$ times the
total remaining mass (after removing the first largest). If this is the case,
we say that there is dominance at level 2. If not, the second largest droplet
is removed and we proceed to do the same check with the third largest one, and
so forth, until only the last droplet remains.
\\
These ideas can be formalized as follows. First all droplets in a given
fragmentation event are ordered by size: $x_1 > x_2 > \ldots > x_n$. Then for
a given value of the ``dominance parameter'' $\gamma$, check whether the
dominance condition
\begin{equation}
  x_k > \gamma \sum_{k}^n x_i
  \label{eq:3}
\end{equation}
is satisfied for some $k<n$. Stop at the smallest value of $k$ that satisfies
(\ref{eq:3}). In other words we will say that there is dominance at level $k_d$ if
condition (\ref{eq:3}) is satisfied at $k=k_d$ but not at $k_d-1,
k_d-2,\ldots,1$. If (\ref{eq:3}) is not satisfied for any $k<n-1$ then we say
that there is no dominance. \\
\begin{figure}[h!]
  \centering
  \psfig{figure=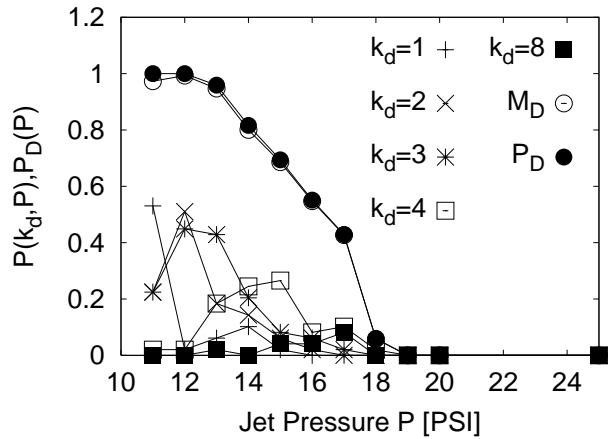,width=6.0cm,angle=270}
  \caption{Probability of dominance $P(k_d,P)$ at $k_d=$ 1 (plusses), 2 (crosses),
    3 (asterisks), 4 (empty squares), and 8 (full squares); mass of dominant
    drops $M_D(P)$ (empty circles), and total dominance probability $P_D(P)$
    (full circles) as a function of pressure $P$.}
  \label{fig:dominance}
\end{figure}
Whenever there is dominance at level $k_d$, the first $k_d$ drops are
substantially larger than all of the remaining ones. This means that the mass
distribution can be separated in two components: one made of $k_d$ massive
drops, the other constituted by one or more (usually many) tiny droplets.  If
on the other hand condition (\ref{eq:3}) is never satisfied, it is said that
there is no dominance and this means that the distribution has no
distinguishable size scale.
\\
In NMF, events where the excited nucleus emits a few small fragments composed
of one or two nuclei, remaining otherwise almost unaltered, are called
\emph{evaporation} events.  Those events where the original nucleus splits
into two large fragments (plus eventually a few much smaller ones) are called
\emph{fission} events, and those for which three large fragments are formed
are called \emph{ternary fission} events.  Within the dominance concept we
have introduced, the NMF classification mentioned above corresponds to
$k_d=1,2$ and $3$ respectively.
\\
In order to apply these ideas to our experimental data, for each pressure $P$
we determine the probability $P_D(k_d,P)$ that there is dominance at level
$k_d$, by averaging over fragmentation events (typically fifty) at pressure
$P$. The resulting dominance probabilities are displayed in
\Fig{fig:dominance} for several values of $k_d$. It is apparent from these
data that evaporation ($k_d=1$) events are predominant at low pressure, next
fission ($k_d=2$) events appear as $P$ is increased, followed by a
predominance of ternary fission($k_d=3$) and so forth. As shown in
\Fig{fig:avgk}, the average dominance level $k_d$ grows on approach to $P=18$
PSI, beyond which there is no longer any detectable dominance.
\\
The sum over $k$ of $P_D(k,P)$ gives probability $D(P)$ that there is
dominance at some level, and is also shown in \Fig{fig:dominance}, together
with the average mass $M_D(P)$ of all dominant droplets. The fact that
$M_D(P)$ is roughly equivalent to $D(P)$ indicates that, whenever there is
dominance, almost the total mass of the system is contained in the $k_d$
dominant drops.
The total dominance probability $D(P)$ goes to zero at around $18 PSI$, in a
way that is consistent with all previously mentioned evidence indicating a
critical point around that pressure. Of course the precise numerical values of
the dominance observables $P_D(k,P)$ depend on the arbitrary choice of the
dominance parameter $\gamma$, but we have checked that the fact that the
dominance probability goes to zero around $18$ PSI holds qualitatively
unaltered for $\gamma$ ranging from $0.8$ to $0.95$.  We also explored a
different definition of dominance, namely replacing (\ref{eq:3}) with
\begin{equation}
  x_k > \gamma x_{k+1}.
  \label{eq:otherdom}
\end{equation}
This means that a comparison is established between each mass-sorted droplet
and the next, stopping at the first level at which the ratio between them is
larger than a given prespecified factor. The use of definition
(\ref{eq:otherdom}) instead of (\ref{eq:3}) did not alter our findings
qualitatively.
\\
We thus feel justified to conclude, based on the data presented in this
section, that there is a \emph{dominance transition} in the neighborhood of
$P_c=18$~PSI. On the low-pressure side of this transition, a few large
(dominant) drops contain most of the original mass, and there is a size-gap
separating these large drops from the accompanying ``mist'' of tiny droplets.
Beyond $P_c$, on the other hand, one has a collection of droplets which
populate homogeneously the whole range of sizes without a systematic size gap.
\section{Discussion}
\label{sec:discussion}
We have presented a statistical analysis of droplet-size distributions
obtained when a liquid drop is broken by a sudden gas jet blow. A flatbed
scanner and image processing software were used to automatize the process of
counting and measuring the resulting fragments, enabling us to collect
statistics for breakdown processes resulting in up to tens of thousands of
tiny droplets. The resulting size-distributions $N(x)$, displayed in
\Fig{fig:Ndens.in.x}, can be fitted by a simplified Fisher-model expression
(\Eqn{eq:14}) and it was found that $N(x)$ behaves like a pure power-law only
at \hbox{$P_c \approx 17\hbox{~PSI}$}, providing the first hint for the
existence of a phase transition there. Further evidence suggesting a phase
transition is given by a deep minimum in the apparent Fisher exponent $\tau$
(\Fig{fig:effexp}), and by a sharp slope discontinuity in the relation between
pressure and surface increase (\Fig{fig:surfincrease}), both of them occurring
at $P_c$.  We argued that it is not possible to identify $P_c$ as a
percolation critical point, as was done in NMF~\cite{RWMMA01,DD-GLTTM05} and
Fracture~\cite{WKHFOS04,WKHBOS05,HWKF06} experiments, because: a)
$\tau^{min}\approx 1$ (\Fig{fig:effexp}) is far from the value $2.3$ that
would be expected for a percolative transition, and b) in our case, the mass
$Z_{max}$ of largest fragment does not become zero at $P_c$, but instead
behaves as $P^{-3/2}$ in the whole pressure range considered (See
\Fig{fig:holianpred}). Low values of the apparent exponent $\tau$ have been
previously reported in NMF experiments\cite{PCTEFA84,PCSFII85,MBCEOP88}.
\\
Furthermore, the behavior of the largest and average fragments, in our
experiments, was found to be consistent with recent predictions of Ashurst and
Holian (AH)~\cite{AHDFB99,AHDSD99}, if the original droplet is assumed to
expand at rate $\eta$ that is proportional to $P$ (\Fig{fig:holianpred}).
\\
Analysis of moment correlations through a scatter plot (see
\Fig{fig:ZmaxvsS2}) reveals a behavior similar to the one already found by
Campi~\cite{CMNB86} in NMF experiments, and which was invoked as evidencing
percolative critical behavior. We showed, however, that the shape of the
scatter plot displayed in \Fig{fig:ZmaxvsS2}, i.e. the existence of a maximum
in the values of $S_2(Z)$, is in no way related to criticality, since it is
there even for a simple, non-critical, partitioning model, as discussed in
\Sec{sec:observ-trans-perc}.
\\
We introduced the concept of ``dominance'' in an attempt to quantify the
observed fact that a few large droplets contain most of the original mass at
low pressures, while no dominant droplets are observed at larger pressures.
The dominance probability, as we defined it, although is dependent upon an
arbitrary parameter $\gamma$, shows a sharp fall (\Fig{fig:dominance}) to zero
at a critical pressure $P_c \approx 18$ PSI, that is roughly independent of
$\gamma$. Of course the precise value of $P_c$ is not interesting by itself,
since it will depend upon many parameters like the volume and viscosity of the
original droplet, the amount of gas contained in the pressurized chamber, the
diameter of the outlet (and thus the exhaust velocity), etc, whose variation
we did not consider in this experiment. Conceptually more interesting is the
fact that, as the energy available to break the droplet increases, a sharp
transition occurs between two breakdown regimes, and that this transition can
be characterized by statistical methods as discussed in \Sec{sec:results}.
\\
Future work could include the use of fast imaging techniques in order to put
these results in the context of what is known about rupture modes in the field
of sprays~\cite{PEUOB87,HFNDD92,JBBBOA99,LREOL01}. The following picture
emerges for isolated droplets subject to a sudden gas blow.  At low excitation
intensity (low Weber number) the prevalent rupture mechanism is vibrational
breakup, whereby collective oscillations break the drop into a small number of
roughly equal-sized droplets.  Upon increasing the breakdown energy, bag
breakup appears, which is characterized by the deformation of the drop into a
torus-shaped rim, which subsequently disintegrates. A variant of this is
called bag-and-stamen breakup, in which a central ``stamen'' is also formed.
Next come shear breakup, where small droplets are continuously stripped off
the rim of the drop, and catastrophic breakup, where strong surface waves
disintegrate the drop violently. The limit Weber number between vibrational
and bag breakup is around 10. At 20-60 shear breakup appears, and at 1000
catastrophic breakup.
\\
Since our experiments were conducted on droplets hanging from a thin glass
tip, the presence of which certainly modifies the gas flow and the ensuing
breakup modes, we do not expect a direct correspondence with the above
classification, although a comparison with the rupture modes of free droplets
would certainly be instructive.
\\
Also interesting is to establish a comparison of our results with recent
analyses of NMF data in the context of phase
transitions~\cite{RWMMA01,DD-GLTTM05}. Common characteristics are: a sharp
minimum in the apparent $\tau$ exponent~\cite{PCTEFA84,MBCEOP88,RWMMA01}, a
strong curvature change in a log-log plot of $N(x)$\cite{BLBCEO95,GFFPTL01},
and the fact that $N(x)$ is, to a good approximation, a pure power law at
criticality~\cite{PCTEFA84,MBCEOP88,BLBCEO95,RWMMA01,GFFPTL01}.
\begin{acknowledgments}
  This work has been partially supported by Conacyt, M\'exico, under Grants
  No.  46709-F and 48783-F.
\end{acknowledgments}

\end{document}